# High brightness electron beam for radiation therapy – A new approach

Wei Gai (盖炜)

Engineering Physics Department, Tsinghua University, Beijing, China

Argonne National Laboratory, Argonne, IL 60439, USA

wg@anl.gov

June 28, 2017

Abstract:  I propose to use high brightness electron beam with 1 – 100 MeV energy as tool to combat tumor/cancerous tissues in deep part of body.  The method is to directly deliver the electron beam to the tumor site via a small tube that connected to a high brightness electron beam accelerator that is commonly available around the world.  Here I gave a basic scheme on the principle, I believe other issues people raises will be solved easily for those who are interested in solving the problems.

Main Text

Charged particle beam has been used for radiation therapy for many decades.  Their effects have been demonstrated and many life saved, and should be further explored 【1】.

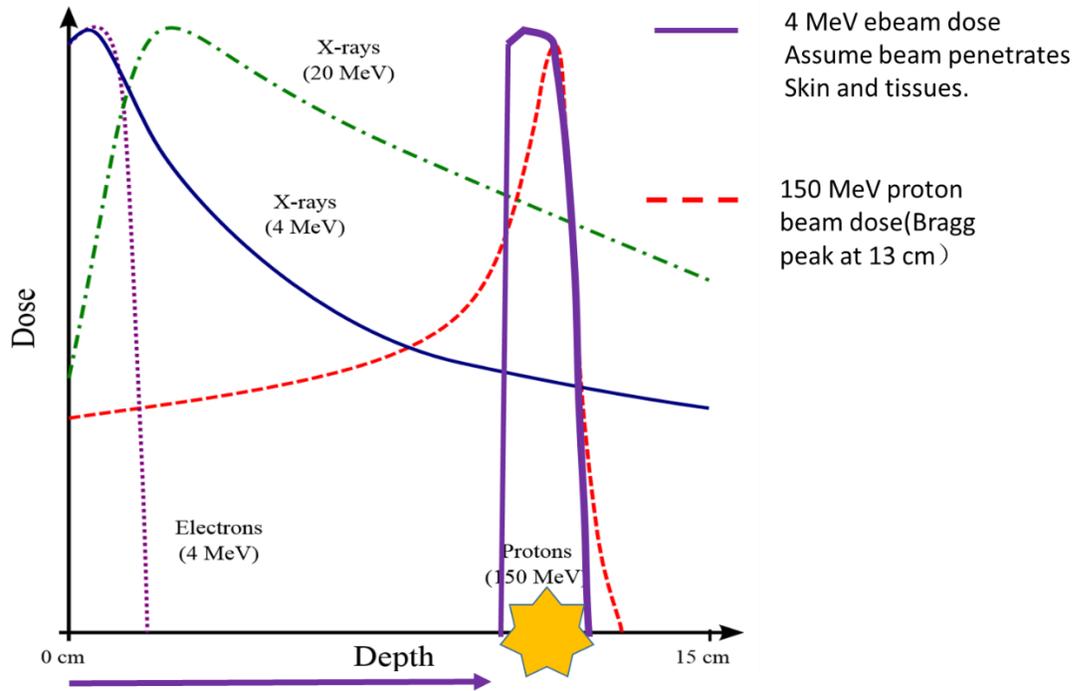

Figure 1 absorbed dose curves for different particle rays (X, e, and p).   Notice the vertical scale is linear. (part from Google Images, Bragg Peak)

As we noted in Figure 1, if there is a tumor at site near 13 cm, when both X-ray and proton beam reach the site, it will deposit large amount doses in healthy tissues.  However, one notice that for 4 MeV electron beam, the absorption curve sharply drops to zero after 1 – 2 cm penetrating into the skin (purple dash line), thus cannot reach to tumor site.   Based on this observation, move of a pristine electron beam directly to the tumor site in 13 cm would have the same effect like treating skin or shallow tumors, as the solid purple curve indicates.  This is like using a Mobetron [ref] but without open surgery.

Based on the curves, we now consider the following scheme for treatment as in Figure 2.  With a high brightness electron beam accelerator producing beam with high directivity like a laser beam, and we could extract the beam and then pass through a small tube that punctuate the skin and healthy tissues to the tumor site.

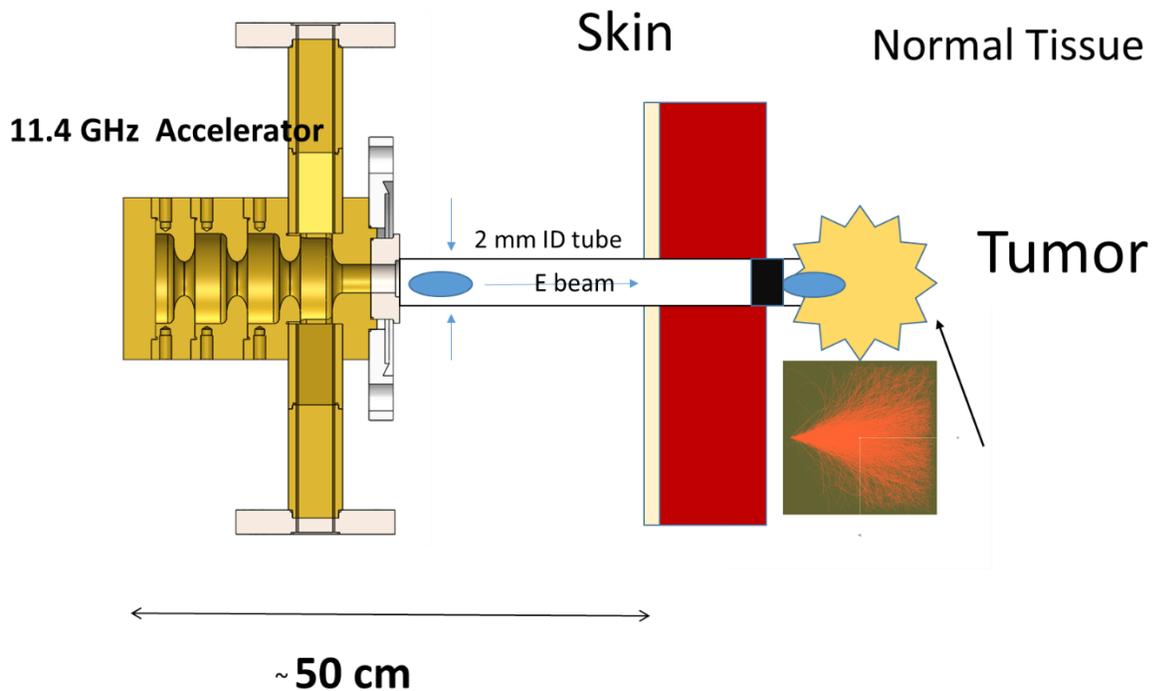

Figure 2. Schematic illustration showing how a beam could reach to the tumor site. A high quality electron beam from cathode, then accelerated via RF cavities to suitable energies (4 MeV, for example), could pass through a small tube with a few mm diameter in tens of cm to reach a tumor site.

One inspiration we had was that a similar procedure called biopsy is commonly performed to identify the cancerous tissues. It surgically inserts a small tube with diameter of a few mm to the tumor site and then extract the tissues for lab tests. One could be imaging a similar method that we insert a small tube that open up a small passage for the electron beam to reach the tumor site with no beam losses long the way that preserve the healthy tissues, then all the radiation dosages will be deposited in the tumor site.

High brightness electron beam technologies have been around for many years (ipac conference proceedings, for example). Here I gave an accelerator injector design for X-band (11.424 GHz) that we are now considering at Tsinghua University, and show that a well-adjusted beam with 200 pico-Coulomb charge from photocathode RF gun can travel in a small vacuum tube for long distance (50 – 100 cm), as shown in Figure 3. At end of tube, one could attach a thin and low Z material (such as Ti or Be) that has little scattering of electron beam. Although this design has not been realized in the Lab., but almost any S-band RF photocathode gun ( < 1mm- mrad normalized emittance and 4 MeV) would be able to do the same.

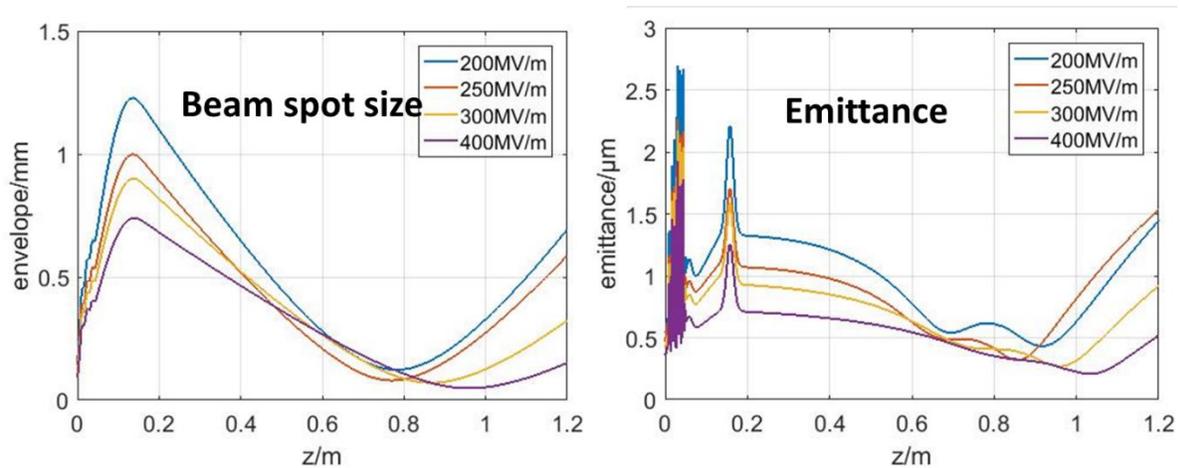

Figure 3 A simulated electron beam envelope as in Figure 2.  Beam produced at Z=0, with accelerating fields ranging from 200 MV/m – 400 MV/m.  However, lower fields would also produce the same kind beam parameters.  The charge here is 200 pC.  Lower charge per pulse would yield even better beam quality.   (accelerator design courtesy L. Zhou, a student of Tsinghua U.)

| Ec [MV/m] | $\varepsilon_x$ [μm] | $\sigma_x$ [mm] | B0 [T] | Beam energy [MeV] |
|---|---|---|---|---|
| 200 | 0.42 | 0.12 | 0.36 | 4.7 |
| 250 | 0.32 | 0.08 | 0.47 | 5.8 |
| 300 | 0.26 | 0.07 | 0.55 | 7.0 |
| 400 | 0.20 | 0.04 |  | 9.3 |

Table 1:  beam parameters from the designed X-band accelerator. (Courtesy L. Zhou)

We now estimate the dosage deliver capability.  A typical accelerator can produce 200 pC charge at ~ 4 MeV at rep-rate of ~ 100 Hz, which contains 80 Joules energy/second.  Giving it deposits all the energy within 1 cm³ volume, it would equate 80 Gy/second doses, that will certainly kill almost any cancerous tissues.   Different scenarios can be scaled from here, such as higher energy beams for larger tumors, and modified beam deliver system for shapes.

Other advantages I could envision here is, although without any elaboration, but can be easily proofed: reducing the radiation shields, no need to generate gamma rays, no need a huge room, a carefully local shielding will be enough; reducing or eliminating complexity of the gantry; reducing the side-effects, only a small incision is needed, no/almost no radiation damage on healthy tissues.

There are could be many other benefits that I missed. But bottom line is based on the physics estimation and principle of the radiation therapy, I am confident the scheme is valid and to be used in treatments soon and many patient's life would be saved and quality of life would be improved.

Acknowledgement:  Although I am not supported by anyone to perform this work, but I have benefited from many discussions with my colleagues at Euclid techlabs, Argonne National Lab., and Tsinghua Univ., particularly, and here I thank them all.  Also, I thank oncologists, Dr. Huijun Xu and Dr. Hao Wu,  in Beijing Hospitals, for their valuable advices on medical demands and issues.